\documentclass[12pt,a4paper]{article}
\usepackage{amsmath}
\usepackage{amssymb}
\usepackage{amscd}

\allowdisplaybreaks

\def\T{{\tilde T}}
\def\M0{{\tilde M}_0}
\def\Mp{{\tilde M}_+}
\def\Mm{{\tilde M}_-}
\def\tm{{\tilde \mu}}
\begin{document}

\begin{titlepage}
\title{
\hfill\parbox{4cm}
{\normalsize YITP-01-69\\{\tt hep-th/0110124}}\\
\vspace{1cm}
Some Properties of String Field Algebra}
\author{
Isao {\sc Kishimoto}\thanks{{\tt
    ikishimo@yukawa.kyoto-u.ac.jp}}
\\[7pt]
{\it Yukawa Institute for Theoretical Physics, Kyoto University,}\\
{\it  Kyoto 606-8502, Japan}
}
\date{\normalsize October, 2001}
\maketitle
\thispagestyle{empty}

\begin{abstract}
\normalsize
We examine string field algebra which is generated by star product in Witten's string field theory including ghost part. We perform calculations using oscillator representation consistently. We construct wedge like states in ghost part and investigate algebras among them.  As a by-product we have obtained some solutions of vacuum string field theory. We also discuss some problems about identity state. We hope these calculations will be useful for further investigation of Witten type string field theory.
\end{abstract}

\end{titlepage}

\section{Introduction}

Recently, string field theory has been discussed again in the context of tachyon condensation\footnote{
	For a review of recent works, see \cite{KO} for example.
} and it is important to get exact solutions of it for reasonable discussion. 
Motivated by development of noncommutative field theory, some techniques are discussed especially to construct analogy of projectors in matter part of open string field theory with Witten type $\star$ product \cite{Witten}.
String field algebra with respect to the $\star$ product has been often discussed only in matter part, but to solve equation of motion of Witten type string field theory, we should consider ghost part more seriously. It is important to develop some techniques in ghost part corresponding to those in matter part and discuss some properties for this purpose.

In this paper, we expand some techniques of string field algebra which was mainly developed in \cite{RSZ}\cite{FO} into ghost part by using oscillator representation consistently.\footnote{
Although there is  half-string formulation which was developed in \cite{HALF}\cite{RSZ}\cite{KOM} (and references therein) and used for concrete calculations of the $\star$ product in matter part, we do not use it because treatment of zero modes in both matter and ghost part is rather subtle.}
We can calculate the $\star$ product analogously in $bc$ ghost nonzero modes by using some relations among Neumann coefficients similar to those in matter part, but we should treat ghost zero modes carefully. Differences of formulas between matter part and ghost part come from ghost zero modes or the balance of ghost number. As a by product of our formulations, we get some classical solutions of vacuum string field theory (VSFT)\footnote{
	For a review, see \cite{VSFT}.
} in the Siegel gauge and resolve some mystery about identity state in the oscillator language.

We hope these calculations will be useful to get exact solutions of original Witten's string field theory \cite{Witten} or analyze VSFT thoroughly.

This paper is organized as follows.
In \S \ref{sec:VER}, we review constructions of Witten's string vertex which gives the definition of $\star$ product between string fields and give some useful relations among Neumann coefficients which we use very often in later concrete calculations.
In \S \ref{sec:ALG}, we investigate algebras among squeezed states. Especially, we define {\it reduced} star product which simplifies formulas and calculations of the $\star$ product in ghost part.
In \S \ref{sec:VSFT}, we get solutions of VSFT by using techniques in previous sections and discuss some issues about identity state.
In \S \ref{sec:DIS}, we discuss some prospects and future problems.
In appendix, we collect our conventions and some useful formulas.

\section{Witten's string vertex \label{sec:VER}}

Here we review constructions of Witten's 3-string vertex including ghost part, and show some relations among Neumann coefficients which we often use for later calculations of the $\star$ product in both matter and ghost part.

There are two ways to construct oscillator representation of Witten's 3-string vertex: one is purely algebraic treatment and the other is Neumann function method. Although the former has advantage to find relations among Neumann coefficients, there is a little nuisance to treat zero modes in ghost part. We relate both techniques by using a construction of \cite{IOS} in which 6-string vertex in matter part leads 3-string vertex in matter and ghost part and show some relations between Neumann coefficients including ghost part.

\subsection{Witten's $N$-string vertex in matter part \label{sec:GJN}}

We can obtain matter $N$-string vertex algebraically.
We follow the procedure of construction in \cite{GJ(I)}.
Actually only $N=6$ case is necessary for our purpose to define the $\star$ product.

The connection condition on Witten type $N$-string vertex $|V_N\rangle$ is
\begin{equation}
\label{eqn:OLX}
X^{(r)}(\sigma)-X^{(r-1)}(\pi-\sigma)=0,\ P^{(r)}(\sigma)+P^{(r-1)}(\pi-\sigma)=0,\ \ 0\leq\sigma\leq{\pi\over2},r=1,\cdots,N.
\end{equation}

Using Fourier transformed coordinate:
\begin{eqnarray}
Q_k(\sigma)&:=&{1\over\sqrt N}\sum_{r=1}^N \omega_N^{kr}X^{(r)}(\sigma)=Q_{k0}+\sqrt2\sum_{n=1}^\infty Q_{kn}\cos n\sigma,\nonumber \\
P_k(\sigma)&:=&{1\over\sqrt N}\sum_{r=1}^N \omega_N^{kr}P^{(r)}(\sigma)={1\over\pi}\left(P_{k0}+\sqrt2\sum_{n=1}^\infty P_{kn}\cos n\sigma\right),\ \ \omega_N=\exp\left({2\pi i\over N}\right),
\end{eqnarray}
this connection condition is rewritten as
\begin{eqnarray}
Q_k(\sigma)&=&\left\{
\begin{array}{lr}
\omega_N^kQ_k(\pi-\sigma)      &    0\leq\sigma\leq{\pi\over2}   \\
\omega_N^{-k}Q_k(\pi-\sigma)   &    {\pi\over2}\leq\sigma\leq\pi  \\
\end{array}
\right.,\nonumber \\
P_k(\sigma)&=&\left\{
\begin{array}{lr}
-\omega_N^kP_k(\pi-\sigma)      &    0\leq\sigma\leq{\pi\over2}   \\
-\omega_N^{-k}P_k(\pi-\sigma)   &    {\pi\over2}\leq\sigma\leq\pi  \\
\end{array}
\right.\ \ (k=1,\cdots,N)
\end{eqnarray}
or
\begin{equation}
(1-Y_k)|Q_k)=(1+Y_k)|P_k)=0,
\end{equation}
where
\begin{eqnarray}
&&Y_k=\cos\left({2\pi k\over N}\right) C +\sin\left({2\pi k\over N}\right)X_{\rm GJ},\ \ C_{nm}=(-1)^n\delta_{n,m}, \ \ n,m\geq0,\nonumber \\
&&X_{\rm GJ}(\sigma,\sigma')=i\left(\theta\left({\pi\over2}-\sigma\right)-\theta\left(\sigma-{\pi\over2}\right)\right)\delta(\sigma+\sigma'-\pi),\nonumber \\
&&X_{\rm GJ}=X_{\rm GJ}^\dagger,X_{\rm GJ}^2=1,{\bar X}_{\rm GJ}=X_{\rm GJ}^T=-X_{\rm GJ}=CX_{\rm GJ}C,\nonumber \\
&&Y_k^\dagger=Y_k,Y_k^2=1,{\bar Y}_k=Y_k^T=Y_{-k}=CY_kC.
\end{eqnarray}
Under the ansatz
\begin{equation}
|V_N\rangle=\mu_N e^{E_N}|0\rangle,\ \ E_N=-{1\over2}\sum_{k=1}^N A_k^\dagger U_k A_{N-k}^\dagger,
\end{equation}
where $\mu_N$ is normalization constant and
using relations of Fourier transformed creation-annihilation operators:
\begin{eqnarray}
&&A_{kn}={1\over\sqrt N}\sum_{r=1}^N \omega_N^{rk} a_n^{(r)},\ \ 
A^\dagger_{kn}={1\over\sqrt N}\sum_{r=1}^N \omega_N^{-rk} a_n^{(r)\dagger},\ \ [A_{kn},A^\dagger_{lm}]=\delta_{k,l}\delta_{n,m}\ \ (n,m\geq0)\nonumber\\
&&Q_{kn}={i\over\sqrt{2n}}\sqrt{2\alpha'}(A_{kn}-A^\dagger_{-kn}),\ \ 
P_{kn}=\sqrt{n\over2}{1\over\sqrt{2\alpha'}}(A_{kn}+A^\dagger_{-kn}),\ \ [Q_{kn},P_{lm}]=i\delta_{k+l,0}\delta_{n,m}\nonumber\\
&&Q_{k0}={i\over2}\sqrt{2\alpha'}(A_{k0}-A^\dagger_{-k0}),\ \ 
P_{kn}={1\over\sqrt{2\alpha'}}(A_{k0}+A^\dagger_{-k0}),\ \ [Q_{k0},P_{l0}]=i\delta_{k+l,0},
\end{eqnarray}
connection condition is rewritten as equations of matrices:
\begin{eqnarray}
&&(1-Y_k)E(1+U_k)=0,\ (1+Y_k)E^{-1}(1-U_k)=0,\nonumber\\
&&(1-Y_k^T)E(1+U_k^T)=0,\ (1+Y_k^T)E^{-1}(1-U_k^T)=0
\end{eqnarray}
where
\begin{equation}
E_{00}={1\over\sqrt2},\ \ E_{nm}={1\over\sqrt n}\delta_{n,m},\ \ E_{n0}=E_{0m}=0,\ \ n,m\geq1.
\end{equation}
Its solution is given by
\begin{eqnarray}
U_k&=&(2-EY_kE^{-1}+E^{-1}Y_kE)(EY_kE^{-1}+E^{-1}Y_kE)^{-1}\nonumber\\
&=&(E^{-1}Y_kE+EY_kE^{-1})^{-1}(2-E^{-1}Y_kE+EY_kE^{-1}),\nonumber\\
&&U_k^T={\bar U_k}=CU_kC=U_{N-k},\ U_k^2=1,\ U_N=C.%,(U_{N\over2}=-C).
\end{eqnarray}
Finally, we obtain $N$-string vertex as
\begin{eqnarray}
\label{eqn:NGJ}
|V_N\rangle&=&\mu_N e^{E_N}|0\rangle,\nonumber\\
E_N&=&-{1\over2}\sum_{r,s=1}^N a^{(r)\dagger}U^{rs}a^{(s)\dagger}=
-{1\over2}\sum_{r,s=1}^N \sum_{n,m\geq0} a^{(r)\dagger}_n U^{rs}_{nm}a^{(s)\dagger}_m,\nonumber\\
U^{rs}&=&{1\over N}\sum_{k=1}^N\omega_N^{k(r-s)}U_k,\ \ U_k=\sum_{t=1}^N\omega_N^{-kt}U^{tN},\nonumber\\
&&U^{r+1s+1}=U^{rs},\ U^{rsT}=U^{sr},\ {\bar U^{rs}}=CU^{sr}C.
\end{eqnarray}

\subsection{Neumann coefficients in matter part \label{sec:NEU}}

We get relations between usual Neumann coefficients and `Gross-Jevicki' coefficients by treating zero modes carefully.\footnote{
This was done for $N=3$ case in \cite{RSZ0102} (Appendix B).
}

By using Neumann function method, we can construct vertex which satisfies connection condition in the following form expressed using momentum for zero modes:
\begin{eqnarray}
\label{eqn:NEU}
|V_N\rangle&=&{\tilde \mu}_N \int d^dp^{(1)}\cdots d^dp^{(N)} (2\pi)^d\delta^d(p^{(1)}+\cdots+p^{(N)})e^{{\tilde E}_N}|0,p\rangle,\nonumber \\
{\tilde E}_N&=&-{1\over2}\sum_{r,s=1}^N \sum_{n.m\geq1}a^{(r)\dagger}_nV^{rs}_{nm}a^{(s)\dagger}_m-\sum_{r,s=1}^N \sum_{n\geq1}p^{(r)}V^{rs}_{0n}a_n^{(s)\dagger}-{1\over2}\sum_{r,s=1}^Np^{(r)}V^{rs}_{00}p^{(s)},\nonumber \\
&&V^{r+1,s+1}_{nm}=V^{rs}_{nm},\ \ V^{rs}_{nm}=V^{sr}_{mn},
\end{eqnarray}
where zero mode basis are given by 
\begin{eqnarray}
|0,p\rangle&=&|0,p^{(1)}\rangle|0,p^{(2)}\rangle\cdots|0,p^{(N)}\rangle,\nonumber \\
|0,p^{(r)}\rangle&=&\left({2\pi\over b}\right)^{-{d\over4}}e^{ip^{(r)}{\hat x}^{(r)}}e^{-{1\over2}(a_0^{\dagger(r)})^2}|0\rangle\nonumber \\
&=&\left({2\pi\over b}\right)^{-{d\over4}}\exp\left(-{b\over4}p^{(r)}p^{(r)}+\sqrt b a_0^{(r)\dagger}p^{(r)}-{1\over2}a_0^{(r)\dagger}a_0^{(r)\dagger}\right)|0\rangle,\ \ b=2\alpha'.
\end{eqnarray}
From momentum conservation, we can redefine as $V^{rs}_{0n}\rightarrow V^{rs}_{0n}+v_n^s$.
We choose $V^{rs}_{0n}$ for simplicity such that
\begin{equation}
\sum_{t,s=1}^N (A^{-1})^{st}V_{0n}^{tr}=0,\ \ \ A^{rs}:={b\over2}\delta^{r,s}+V_{00}^{rs}.
\end{equation}

To relate coefficients $V^{rs}_{nm}$ with $U^{rs}_{nm}$, we rewrite eq.(\ref{eqn:NEU}) into the form eq.(\ref{eqn:NGJ}) by performing Gaussian integral with respect to momentum. We get the relations between $U^{rs}$ and $V^{rs}$:
\begin{eqnarray}
U_{nm}&=&V_{nm}-V_{n0}A^{-1}V_{0m}, \ \ n,m\geq1,\nonumber \\
U_{0n}&=&\sqrt bA^{-1}V_{0n},\ \ n\geq1,\nonumber\\
U_{00}^{rs}&=&b\left({\sum_t\left(A^{-1}\right)^{rt}\sum_u\left(A^{-1}\right)^{su}\over\sum_{r',s'}\left(A^{-1}\right)^{r's'}}-\left(A^{-1}\right)^{rs}\right)+\delta^{r,s}.
\end{eqnarray}
Conversely, we get
\begin{eqnarray}
V_{00}^{rs}&=&{b\over2N}\sum_{k=1}^{N-1}\omega_N^{k(r-s)} {1+(U_k)_{00}\over1-(U_k)_{00}}
={b\over2N}\sum_{k=1}^{N-1}\cos\left({2\pi k(r-s)\over N}\right) {1+(U_k)_{00}\over1-(U_k)_{00}},\nonumber \\
V_{0n}^{rs}&=&V_{n0}^{sr}={\sqrt b\over N}\sum_{k=1}^{N-1}\omega_N^{k(r-s)}{1\over1-(U_k)_{00}}(U_k)_{0n}, \nonumber \\
V_{nm}^{rs}&=&{1\over N}C_{nm}+{1\over N}\sum_{k=1}^{N-1}\omega_N^{k(r-s)}\left((U_k)_{nm}+(U_k)_{n0}{1\over1-(U_k)_{00}}(U_k)_{0m}\right),
\end{eqnarray}
where the first line corresponds to the convention:\footnote{
	This ambiguity comes from momentum preservation.
}
\begin{equation}
2V_{00}^{rs}=\sum_{t=1}^N\sum_{n\geq1}V_{0n}^{rt}V_{0n}^{st},\ \ \sum_{t,u=1}^N V_{00}^{tu}=0.
\end{equation}
We can also rewrite $V^{rs}_{nm}\ (n,m\geq1)$ like $U^{rs}_{nm}\ (n,m\geq0)$ in (\ref{eqn:NGJ}) :
\begin{eqnarray}
V_{nm}^{rs}&=&{1\over N}\sum_{k=1}^N\omega_N^{k(r-s)}({\tilde U}_k)_{nm},\ \ \  n,m\geq1, \nonumber \\
({\tilde U}_k)_{nm}&:=&(U_k)_{nm}+(U_k)_{n0}{1\over1-(U_k)_{00}}(U_k)_{0m},\ \ \ k=1,\cdots,N-1, \nonumber \\
{\tilde U}_N&:=&C, \nonumber \\
&&{\bar {\tilde U}}_k={\tilde U}_k^T=C{\tilde U}_kC={\tilde U}_{N-k},\ {\tilde U}_k^2=1.
\end{eqnarray}

In this convention for $V_{0n}^{rs}$, we can show
\begin{equation}
\label{eqn:VV}
\sum_{t=1}^N \sum_{n=1}^\infty V_{mn}^{rt}V_{n0}^{ts}=V_{m0}^{rs},\ \ \ 
\sum_{t=1}^N \sum_{m=1}^\infty V_{nm}^{rt}V_{ml}^{ts}=\delta_{n,l}\delta_{r,s}.
\end{equation}

Finally, we get the relation between normalizations of $|V_N\rangle$ :
\begin{equation}
\mu_N={\tilde \mu}_N \left({(2\pi)^{N+1}\over\det_{r,s}A^{rs} \sum_{r,s=1}^N (A^{-1})^{rs}}\right)^{d\over2}\left({2\pi\over b}\right)^{-{N d\over4}}.
\end{equation}
%Note this does not depend on the convention of $V_{00}^{rs}$.

\subsection{From 6-string vertex to 3-string vertex including ghost part}

We can construct 3-string vertex from matter 6-string vertex including ghost part.\footnote{
	We use some results in the reference \cite{IOS}.
}

The connection condition on Witten type 3-string vertex in ghost part is
\begin{equation}
c^{\pm(r)}(\sigma)+c^{\pm(r-1)}(\pi-\sigma)=0,\ \ b^{\pm(r)}(\sigma)-b^{\pm(r-1)}(\pi-\sigma)=0,\ 0\leq\sigma\leq{\pi\over2},\ \ \ r=1,2,3,
\end{equation}
and in matter part are the same as (\ref{eqn:OLX}) in $N=3$ case.
The 3-string vertex which satisfies these conditions is given as follows \cite{IOS}:
\begin{eqnarray}
\label{eqn:3SV}
|V_3\rangle&=&{\tilde \mu}_3 \int d^dp^{(1)}d^dp^{(2)}d^dp^{(3)} (2\pi)^d\delta^d(p^{(1)}+p^{(2)}+p^{(3)})e^{{\tilde E}_3}|0,p\rangle,\nonumber\\
{\tilde E}_3&=&-{1\over2}\sum_{r,s=1}^3 \sum_{n.m\geq1}a^{(r)\dagger}_nV^{rs}_{nm}a^{(s)\dagger}_m-\sum_{r,s=1}^3 \sum_{n\geq1}p^{(r)}V^{rs}_{0n}a_n^{(r)\dagger}-{1\over2}\sum_{r,s=1}^3p^{(r)}V^{rs}_{00}p^{(s)}\nonumber \\
&&-\sum_{r,s=1}^3 \sum_{n\geq1,m\geq0}c_{-n}^{(r)}X^{rs}_{nm}b_{-m}^{(s)},\nonumber \\
|0,p\rangle&=&|0,p^{(1)}\rangle|0,p^{(2)}\rangle|0,p^{(3)}\rangle,\nonumber \\
&&b^{(i)}_n|0,p^{(i)}\rangle=0,\ \ n\geq1, \ \ c^{(i)}_m|0,p^{(i)}\rangle=0,\ \ m\geq0,
\end{eqnarray}
where these Neumann coefficients $V^{rs}_{nm},X^{rs}_{nm}$ are given by using 6-string vertex in matter part as follows.
In $N=6$ case in matter part, denoting as
\begin{equation}
U_1=:{\tilde U},\ \ U_2=:U,\ \ U_3=-C,\ \ U_4={\bar U},\ \ U_5={\bar {\tilde U}},\ \ U_6=C,
\end{equation}
in the notation of \S \ref{sec:GJN}, and Neumann coefficients in \S \ref{sec:NEU} as $V^{(6)}{}^{rs}_{nm}$, we get the relations :
\begin{eqnarray}
\label{eqn:VX3}
V_{nm}^{rs}&=&(V^{(6)rs}+V^{(6)rs+3})_{nm}={1\over3}\left(C+\omega^{r-s}U_M+\omega^{s-r}{\bar U}_M\right)_{nm},\nonumber\\
{1\over\sqrt m}X^{rs}_{mn}\sqrt n&=&(-1)^{r+s}(V^{(6)rs}-V^{(6)rs+3})_{mn}={1\over3}\left(-C+\omega^{s-r}U_G+\omega^{r-s}{\bar U}_G\right)_{mn},
\end{eqnarray}
where\footnote{
We denote the matrix $(-1)^n\delta_{n,m},\ n,m\geq1$ as $C$ again which leads to no confusion.
}
\begin{eqnarray}
&&U_{Mmn}=U_{mn}+U_{m0}{1\over1-U_{00}}U_{0n},\ \ U_{Gmn}={\tilde U}_{mn}+{\tilde U}_{m0}{1\over1-{\tilde U}_{00}}{\tilde U}_{0n},\ \ m,n\geq1, \nonumber\\
&&{\bar U}_M=U_M^T=CU_MC,\ \ {\bar U}_G=U_G^T=CU_GC,\ \ U_M^2=1,\ \ U_G^2=1,\  \ \ \omega=\exp\left({2\pi i\over3}\right).
\end{eqnarray}
We can show
\begin{equation}
\sum_{t=1}^3\sum_{l\geq1}V_{nl}^{rt}V_{lm}^{ts}=\delta_{nm}\delta^{rs},\ \ \ 
\sum_{t=1}^3\sum_{l\geq1}X_{nl}^{rt}X_{lm}^{ts}=\delta_{nm}\delta^{rs},
\end{equation}
which correspond to eq.(\ref{eqn:VV}). For the matrices
\begin{equation}
M_0:=CV^{rr},\ \ M_{\pm}:=CV^{rr\pm1},\ \ {\tilde M}_0:=-CX^{rr},\ \ {\tilde M}_{\pm}:=-CX^{rr\pm1},
\end{equation}
where these indices run from $m=1$ to $\infty$, we can show relations :
\begin{eqnarray}
\label{eqn:MMT}
&&CM_0=M_0C,\ CM_+=M_-C,\ \ C\M0 =\M0 C,\ C\Mp =\Mm C,\nonumber \\
&&[M_0,M_{\pm}]=[M_+,M_-]=0,\ \ [{\tilde M}_0,{\tilde M}_{\pm}]=[{\tilde M}_+,{\tilde M}_-]=0,\nonumber \\
&&M_0+M_++M_-=1,\ {\tilde M}_0+{\tilde M}_++{\tilde M}_-=1,\nonumber \\
&&M_+M_-=M_0^2-M_0,\ \ {\tilde M}_+{\tilde M}_-={\tilde M}_0^2-{\tilde M}_0,\nonumber \\
&&M_0^2+M_+^2+M_-^2=1,\ \ \M0^2+\Mp^2+\Mm^2=1,\nonumber \\
&&M_0M_+ +M_+M_- +M_- M_0=0,\ \ \M0 \Mp +\Mp \Mm +\Mm \M0=0, \nonumber \\
&&M_\pm^2-M_\pm=M_0M_\mp ,\ \ {\tilde M}_\pm^2-{\tilde M}_\pm=\M0 {\tilde M}_\mp.
\end{eqnarray}
Neumann coefficients which have zero index are given by
\begin{eqnarray}
V_{0n}^{rs}&=&V_{n0}^{sr}:=(V^{(6)rs}+V^{(6)rs+3})_{0n}={\sqrt b\over3(1-U_{00})}\left(\omega^{r-s}U_{0n}+\omega^{s-r}{\bar U}_{0n}\right), \nonumber \\
V_{00}^{rs}&:=&(V^{(6)rs}+V^{(6)rs+3})_{00}={b\over3}\cos\left({2\pi(r-s)\over3}\right){1+U_{00}\over1-U_{00}}, \nonumber \\
X_{n0}^{rs}&:=&(-1)^{r+s}\sqrt n(V^{(6)rs}-V^{(6)rs+3})_{n0}
={\sqrt{bn}\over3(1-{\tilde U}_{00})}\left(\omega^{r-s}{\tilde U}_{0n}+\omega^{s-r}{\bar {\tilde U}}_{0n}\right).
\end{eqnarray}
Especially, ambiguity which caused by redefinitions of $V^{(6)}{}^{rs}_{0n}$ is canceled in the formula of $X^{rs}_{n0}$.
For these matrices, we can show some relations\footnote{
We use the vector notation as $V^{rs}_{n0}=(V^r{}^s_0)_n,\ X^{rs}_{n0}=(X^r{}^s_0)_n$.
} :
\begin{eqnarray}
\label{eqn:VXTT}
&&\sum_{t=1}^3\sum_{l=1}^\infty V^{rt}_{nl}V_{l0}^{ts}=V_{n0}^{rs},\ \ 
\sum_{t=1}^3\sum_{l=1}^\infty V^{rt}_{0l}V_{l0}^{ts}=2V_{00}^{rs},\ \ 
\sum_{t=1}^3\sum_{l=1}^\infty X^{rt}_{nl}X^{ts}_{l0}=X_{n0}^{rs}.\nonumber \\
&& CV^r{}^s_0=V^s{}^r_0,\ \ \sum_{t=1}^3 V^t{}^s_0=\sum_{t=1}^3V^r{}^t_0=0,\ \ 
CX^r{}^s_0=X^s{}^r_0,\ \ \sum_{t=1}^3 X^t{}^s_0=\sum_{t=1}^3X^r{}^t_0=0,\nonumber \\
&&V^2{}^1_0={3M_+ -2\over1+3M_0}V^1{}^1_0,\ \ V^3{}^1_0={3M_- -2\over1+3M_0}V^1{}^1_0,\nonumber \\
&&X^2{}^1_0=-{\Mp \over1-\M0 }X^1{}^1_0,\ \ X^3{}^1_0=-{\Mm \over1-\M0 }X^1{}^1_0,\ \ 
X^r{}^s_0=(\delta^{rs}+X^{rs}){1\over1+X^{11}}X^1{}^1_0.
\end{eqnarray}
Eqs.(\ref{eqn:MMT})(\ref{eqn:VXTT}) are very useful formulas for concrete calculations of the $\star$ product in the following sections.

For completeness we write down 3-string vertex in oscillator representation:
\begin{eqnarray}
|V_3\rangle&=& \mu_3 e^{E_3}|0\rangle,\nonumber \\
E_3 &=& -{1\over2}\sum_{r,s=1}^3\sum_{n,m\geq0}a_n^{(r)\dagger}U_{nm}^{'rs}a_m^{(s)\dagger}-\sum_{r,s=1}^3 \sum_{n\geq1,m\geq0}c_{-n}^{(r)}X^{rs}_{nm}b_{-m}^{(s)}, \nonumber \\
U^{'rs}_{nm}&=&(U^{rs}+U^{rs+3})_{nm}={1\over3}(C+\omega^{r-s}U+\omega^{s-r}{\bar U})_{nm},\ \ n,m\geq0, \nonumber \\
\mu_3&=&{\tilde \mu}_3\left({2^4(2\pi)^3\over3b^2} \left({1-U_{00}\over1-{1\over3}U_{00}}\right)^2\right)^{d\over2}\left({2\pi\over b}\right)^{-{3d\over4}},\nonumber \\
|0\rangle&=&|0\rangle_1|0\rangle_2|0\rangle_3,\ \ |0\rangle_r=|0\rangle_{M(r)} |+\rangle_{G(r)}, \nonumber \\
&&|+\rangle_{G(r)}=c_0^{(r)}c_1^{(r)}|\Omega\rangle_{G(r)},\ \ 
a^{(r)}_n|0\rangle_{M(r)}=0,\ n\geq0
\end{eqnarray}
where $|\Omega\rangle_G$ is conformal vacuum. This $U^{'rs}_{nm}$ corresponds to (\ref{eqn:NGJ}) of the $N=3$ case in \S \ref{sec:GJN}.\\

Note that $|V_3\rangle$ has cyclic symmetry:
\begin{equation}
|V_3\rangle:=|1,2,3\rangle=|2,3,1\rangle=|3,1,2\rangle.
\end{equation}

\subsection{Reflector and identity state}

In matter part, reflector and identity state correspond to $N=2$ and $N=1$ case in \S \ref{sec:GJN}, but these are rather complicated in ghost part. Here we write down these formulas explicitly in both momentum representation and oscillator representation for matter zero mode.

\subsubsection{Reflector}

The reflector $|V_2\rangle$ is defined as
\begin{eqnarray}
\label{eqn:RFK}
|V_2\rangle&=&|1,2\rangle=\int d^dp^{(1)}d^dp^{(2)}\delta^d(p^{(1)}+p^{(2)})\delta(b^{(1)}_0-b^{(2)}_0)e^{{\tilde E}_2}|0,p\rangle=(b^{(1)}_0-b^{(2)}_0)e^{E_2}|0\rangle, \nonumber \\
{\tilde E}_2&=&-\sum_{n,m\geq1}a^{\dagger(1)}_nC_{nm}a^{\dagger(2)}_m+\sum_{n,m\geq1}(c^{(1)}_{-n}C_{nm}b^{(2)}_{-m}+c^{(2)}_{-n}C_{nm}b^{(1)}_{-m}), \nonumber \\
E_2&=&-\sum_{n,m\geq0}a^{\dagger(1)}_nC_{nm}a^{\dagger(2)}_m+\sum_{n,m\geq1}(c^{(1)}_{-n}C_{nm}b^{(2)}_{-m}+c^{(2)}_{-n}C_{nm}b^{(1)}_{-m}).
\end{eqnarray}
This is antisymmetric: $|1,2\rangle=-|2,1\rangle$.
The bra state of the reflector which gives bra state from ket state generally is
\begin{eqnarray}
\label{eqn:RFB}
\langle V_2|&=&\langle1,2|=\int d^dp^{(1)}d^dp^{(2)}\langle0,p|e^{{\tilde E}'_2}\delta^d(p^{(1)}+p^{(2)})\delta(c^{(1)}_0+c^{(2)}_0)=\langle0|e^{E'_2}(c^{(1)}_0+c^{(2)}_0),\nonumber \\
{\tilde E}'_2&=&-\sum_{n,m\geq1}a^{(1)}_nC_{nm}a^{(2)}_m-\sum_{n,m\geq1}(c^{(1)}_nC_{nm}b^{(2)}_m+c^{(2)}_nC_{nm}b^{(1)}_m), \nonumber \\
E'_2&=&-\sum_{n,m\geq0}a^{(1)}_nC_{nm}a^{(2)}_m-\sum_{n,m\geq1}(c^{(1)}_nC_{nm}b^{(2)}_m+c^{(2)}_nC_{nm}b^{(1)}_m),\nonumber \\
&&\langle0,p|={}_1\langle0,p^{(1)}| {}_2\langle0,p^{(2)}|.
\end{eqnarray}
Note that this is symmetric: $\langle1,2|=\langle2,1|$.
Here we take bra zero mode basis as
\begin{eqnarray}
\langle0,p|&=&\left({2\pi\over b}\right)^{-{d\over4}}\langle0|e^{-{1\over2}a_0a_0}e^{-ip{\hat x}}
=\left({2\pi\over b}\right)^{-{d\over4}}\langle0|e^{-{1\over2}a_0a_0+\sqrt b p a_0-{b\over4}pp},\nonumber \\
\langle0|&=& {}_G\langle{\tilde +}| {}_M\langle0|,\ \ {}_G\langle{\tilde +}|= {}_G\langle\Omega|c_{-1},\ \  {}_M\langle0|a^\dagger_n=0,\ \ n\geq0,
\end{eqnarray}
where ${}_G\langle\Omega|$ is conformal vacuum.
We define normalization as follows:
\begin{eqnarray}
&& {}_M\langle0|0\rangle_M=1,\ \ {}_G\langle{\tilde +}|+\rangle_G= {}_G\langle\Omega|c_{-1}c_0c_1|\Omega\rangle_G=1,\ \ \langle0|0\rangle=1,\ \ \langle0,p|0,p'\rangle=\delta^d(p-p'),\nonumber \\
&&\langle1,2|2,3\rangle=1_{31},\ \ 1_{31}|A\rangle_1=|A\rangle_3 ,\ \ \forall|A\rangle.
\end{eqnarray}

\subsubsection{Identity state}

The identity state is defined in \cite{GJ(I)} as
\begin{eqnarray}
\label{eqn:IKE}
|I\rangle&=&|V_1\rangle={1\over4i}b^+\left({\pi\over2}\right)b^-\left({\pi\over2}\right)\exp\left(\sum_{n\geq1}(-1)^n\left(-{1\over2}a^\dagger_na^\dagger_n+ c_{-n}b_{-n}\right)\right)|0\rangle\nonumber \\
&=&{\tilde \mu}_1(2\pi)^d e^{{\tilde E}_1}b_{-1}b_0|0,0\rangle=\mu_1e^{E_1}b_{-1}b_0|0\rangle,\nonumber \\
{\tilde E}_1&=& -{1\over2}\sum_{n,m\geq1}a^\dagger_nC_{nm}a^\dagger_m+\sum_{n\geq2}(-1)^n c_{-n}b_{-n}\nonumber \\
&&-2 c_0\sum_{n\geq1}(-1)^n b_{-2n}-(c_1-c_{-1})\sum_{n\geq1}(-1)^nb_{-(2n+1)},\nonumber \\
E_1&=&-{1\over2}\sum_{n,m\geq0}a^\dagger_nC_{nm}a^\dagger_m+\sum_{n\geq2}(-1)^n c_{-n}b_{-n} \nonumber \\
&&-2 c_0\sum_{n\geq1}(-1)^n b_{-2n}-(c_1-c_{-1})\sum_{n\geq1}(-1)^nb_{-(2n+1)},\nonumber \\
\mu_1&=&{\tilde \mu}_1(2\pi)^d\left({2\pi\over b}\right)^{-{d\over4}},\ \ \ b_{-1}b_0|0\rangle=|0\rangle_M |\Omega\rangle_G.
\end{eqnarray}

The corresponding bra state (or integration of string field) is\footnote{
Note that this formula coincides with one given by LPP formulation \cite{LPP} which is based on CFT:
\begin{eqnarray}
{}_{\rm LPP}\langle I|&=&\mu_1{}_M\langle0|{}_G\langle\Omega|c_{-1}c_0c_1\int_{\zeta_1\zeta_0\zeta_{-1}}\exp\biggl({1\over2}\sum_{n,m\geq1}\alpha_nN_{nm}\alpha_m\nonumber \\
&&+\sum_{n\geq2,m\geq-1}c_n{\tilde N}_{nm}b_m-\sum_{i=\pm1,0,m\geq1}\zeta_iM_{im}b_m\biggr),\nonumber \\
N_{nm}&=&{1\over nm}\oint {dz\over2\pi i}z^{-n}f'(z)\oint {dw\over2\pi i}w^{-m}f'(w){1\over(f(z)-f(w))^2},\nonumber \\
{\tilde N}_{nm}&=&\oint {dz\over2\pi i}z^{-n+1}(f'(z))^2\oint {dw\over2\pi i}w^{-m-2}(f'(w))^{-1}{-1\over f(z)-f(w)},\nonumber \\
M_{im}&=&\oint {dz\over2\pi i}z^{-m-2}(f'(z))^{-1}(f(z))^{i+1}
\end{eqnarray}
where the map $f(z)$ is defined in \cite{RZ}: $f(z)={2z\over1-z^2}$.
}
\begin{eqnarray}
\label{eqn:IBR}
{}_1\langle I|&:=& {}_{12}\langle V_2|I\rangle_2, \ \ \langle I|={\tilde \mu}_1\langle0,0|b_1e^{{\tilde E}'_1}=\mu_1\langle0|b_1e^{E'_1}, \nonumber \\
{\tilde E}'_1 &=& -{1\over2}\sum_{n,m\geq1} a_n C_{nm}a_m - \sum_{n\geq2}(-1)^n c_n b_n\nonumber \\
&&+ 2 c_0 \sum_{n\geq1}(-1)^n b_{2n} - (c_1-c_{-1})\sum_{n\geq1}(-1)^n b_{2 n+1},\nonumber \\
E'_1 &=& -{1\over2}\sum_{n,m\geq0} a_n C_{nm} a_m - \sum_{n\geq2}(-1)^n c_n b_n\nonumber \\
&&+ 2 c_0 \sum_{n\geq1}(-1)^n b_{2n} - (c_1-c_{-1})\sum_{n\geq1}(-1)^n b_{2 n+1},\ \ \langle0|b_1= {}_M\langle0| {}_G\langle\Omega|.
\end{eqnarray}

\subsection{Witten's $\star$ product and identity state \label{sec:WITI}}

The Witten's $\star$ product of string field is defined by using 3-string vertex (\ref{eqn:3SV}) as :
\begin{equation}
\label{eqn:STR}
|A\star B\rangle_1:= {}_2\langle A| {}_3\langle B|1,2,3\rangle=\langle2,4|A\rangle_4 \langle3,5|B\rangle_5 |1,2,3\rangle.
\end{equation}

We call $\Psi_{\rm id.}$ ``{\it identity}''  with respect to the $\star$ product if $A\star\Psi_{\rm id.}=\Psi_{\rm id.}\star A=A$ for any string field $A$. This condition can be rewritten as ${}_3\langle\Psi_{\rm id.}|1,2,3\rangle=|1,2\rangle$ in our notation.
We can check whether identity state $|I\rangle$ (\ref{eqn:IKE}) is {\it identity}  or not by straightforward calculation in our oscillator representation:
\begin{eqnarray}
\label{eqn:IV2I}
{}_3\langle I|1,2,3\rangle&=&\mu_1\mu_3\left(\det(1-M_0)\right)^{-{d\over2}}\det(1-{\tilde M}_0)|1,2\rangle_M |1,2\rangle'_G, \nonumber \\
|1,2\rangle_M &=& \exp\left(-\sum_{n,m\geq0}a^{\dagger(1)}_nC_{nm}a^{\dagger(2)}_m\right)|0\rangle_{M12}, \nonumber \\
|1,2\rangle'_G&=&(1-2[(1-{\tilde M}_0)^{-1}X^1{}^1_0]_{\cal E})\cdot \nonumber \\
&&\cdot \left([(1-{\tilde M}_0)^{-1}{X^2}{}^1_0]_{\cal O}(b^{(1)}_0-b^{(2)}_0)-[(1-{\tilde M}_0)^{-1}({\tilde M}_+ b^{\dagger(1)}+{\tilde M}_- b^{\dagger(2)})]_{\cal O}\right) \cdot \nonumber \\
&&\cdot \exp\left(\sum_{n,m\geq1}(c^{(1)}_{-n}C_{nm}b^{(2)}_{-m}+c^{(2)}_{-n}C_{nm}b^{(1)}_{-m})\right)e^{\Delta E}|+\rangle_{G12},\nonumber \\
\Delta E&=&\sum_{r,s=1,2}c^{\dagger(r)}(X^{r3}(1-{\tilde M}_0)^{-1} CX^3{}^s_0-X^r{}^s_0)b^{(s)}_0\nonumber \\
&&-2\sum_{r,s=1,2}c^{\dagger(r)}(X^r{}^3_0 -X^{r3}(1-{\tilde M}_0)^{-1}X^1{}^1_0)(1-2[(1-{\tilde M}_0)^{-1}X^1{}^1_0]_{\cal E})^{-1}\cdot \nonumber \\
&&\cdot [(1-{\tilde M}_0)^{-1}C(X^3{}^s_0b^{(s)}_0+X^{3s}b^{\dagger(s)})]_{\cal E} \nonumber \\
&=&-(c^{\dagger(1)}-c^{\dagger(2)}){1\over1-\M0 }X^1{}^1_0(b^{(1)}_0-b^{(2)}_0),
\end{eqnarray}
where we use the notation
\begin{equation}
\label{eqn:OESUM}
[\ ]_{\cal E}:=\sum_{n=1}^\infty(-1)^n[\ ]_{2n},\ \ [\ ]_{\cal O}:=\sum_{n=0}^\infty(-1)^n[\ ]_{2n+1},\ \ b^\dagger_n=b_{-n},c^\dagger_n=c_{-n},\ n\geq1.
\end{equation}
We can rewrite $|1,2\rangle'_G$ as
\begin{eqnarray}
\label{eqn:IGP}
&&|1,2\rangle'_G=(1-2[(1-{\tilde M}_0)^{-1}X^1{}^1_0]_{\cal E})\cdot \nonumber \\
&&\cdot \biggl(\left[{1\over 1-{\tilde M}_0}{X^2}{}^1_0-(c^{\dagger(1)}-c^{\dagger(2)}){1\over1-\M0 }X^1{}^1_0\cdot {1\over1-\M0}(\Mp b^{\dagger(1)}+\Mm b^{\dagger(2)})\right]_{\cal O}|1,2\rangle_G\nonumber \\
&&- \left[{1\over1-\M0 }(\Mp b^{\dagger(1)}+\Mm b^{\dagger(2)})\right]_{\cal O}|V_2^r\rangle_{12}\biggr),\nonumber \\
&&|V^r_2\rangle_{12}:=e^{\sum_{n,m\geq1}(c^{(1)}_{-n}C_{nm}b^{(2)}_{-m}+c^{(2)}_{-n}C_{nm}b^{(1)}_{-m})}|+\rangle_{G12},\ \ |1,2\rangle_G=(b^{(1)}_0-b^{(2)}_0)|V_2^r\rangle_{12}.
\end{eqnarray}
This shows that the identity state $|I\rangle$ is {\it not} identity with respect to $\star$ product although $|I\rangle$ is identity if it is restricted only in matter part.

\section{String field algebra \label{sec:ALG}}

In this section we examine string field algebra of  the $\star$ product between squeezed states in both matter and ghost part by using explicit formulas in previous section. 

\subsection{Matter part}

In this subsection, we restrict calculations in matter part. We consider only zero momentum sector for simplicity.
If we include matter zero modes in oscillator representation, we can perform similar calculations  because relations among Neumann coefficients are the same.
Note that $U^{rs}_{nm}\ n,m\geq0$ (\ref{eqn:NGJ}) in $N=3$ case of \S \ref{sec:GJN} and $V^{rs}_{nm}\ n,m\geq1$ (\ref{eqn:VX3}) satisfy the same relations such as those in (\ref{eqn:MMT}).
This algebra was obtained in \cite{RSZ}\cite{FO} essentially, but we write down string field algebra for comparison with that in ghost part which will be discussed in the following subsection.

\subsubsection{Wedge state}

We define squeezed state $|n_\beta\rangle$ with parameter $\beta$ which correspond to wedge  state as :
\begin{equation}
\label{eqn:MSQ}
|n_\beta\rangle:=e^{\beta a^\dagger}|n\rangle=\mu_n \exp\left(\beta a^\dagger-{1\over2}a^\dagger CT_na^\dagger\right)|0\rangle 
\end{equation}
where $|n\rangle$ is given by the state which is obtained by taking $\star$ product $n-1$ times with a particular squeezed states $|2\rangle$:
\begin{equation}
\label{eqn:NMD}
|n\rangle:=\left(|2\rangle\right)_\star^{n-1},\ \ |2\rangle=\mu_2 e^{-{1\over2}a^\dagger CT_2a^\dagger}|0\rangle.
\end{equation}
For simplicity, we take a matrix $T_2$ which satisfies
\begin{equation}
CT_2=T_2C,\ T_2^T=T_2,\ \ [M_0,T_2]=0,\ \ T_2\not=1.
\end{equation}
then $T_n,\mu_n$ in eq.(\ref{eqn:MSQ}) are given by
\begin{eqnarray}
\label{eqn:TMN}
T_n&=&{T(1-T_2T)^{n-1}+(T_2-T)^{n-1}\over(1-T_2T)^{n-1}+T(T_2-T)^{n-1}},\nonumber \\
\mu_n&=&\mu_2\left(\mu_2\mu_3^M \det{}^{-{d\over2}}\left({1-T\over1-T+T^2}\right)\right)^{n-2}\det{}^{d\over2}\left({1-T^2\over(1-T_2T)^{n-1}+T(T_2-T)^{n-1}}\right),\nonumber \\
M_0&=&{T\over1-T+T^2},
\end{eqnarray}
where $\mu_3^M$ is normalization factor of 3-string vertex in matter part and
 the matrix $T$ is expressed with $M_0$ by solving the quadratic equation : $M_0T^2-(1+M_0)T+M_0=0$.
These formulas are obtained by solving recurrence equation with respect to $n$ \cite{FO}. Note that $|2\rangle$ is Fock vacuum $|0\rangle$ if and only if  $T_2=0$.

We can show the $\star$ product formula by calculating straightforwardly :
\begin{equation}
\label{eqn:MST}
|n_{\beta_1}\star m_{\beta_2}\rangle=\exp\left(-{\cal C}_{n_{\beta_1},m_{\beta_2}}\right)|(n+m-1)_{\beta_1\rho_{1(n,m)}+\beta_2\rho_{2(n,m)}}\rangle,
\end{equation}
where
\begin{eqnarray}
\label{eqn:MSTP}
&&{\cal C}_{n_{\beta_1},m_{\beta_2}}={1\over2}(\beta_1,\beta_2){C\over T_{n,m}}
\left(
\begin{array}{cc}
M_0(1-T_m)  & M_-  \\
M_+     & M_0(1-T_n)
\end{array}
\right)
\left(
\begin{array}{c}
\beta_1^T     \\
\beta_2^T     
\end{array}
\right)={\cal C}_{m_{\beta_2C},n_{\beta_1C}},\nonumber \\
&&\rho_{1(n,m)}={M_-+M_+T_m\over T_{n,m}},\ \ \rho_{2(n,m)}={M_++M_-T_n\over T_{n,m}},\ \ C\rho_{1(n,m)}=\rho_{2(m,n)}C,\nonumber \\
&&T_{n,m}={(1+T)(1-T)^2\over1-T+T^2} {(1-T_2T)^{n+m-2}+T(T_2-T)^{n+m-2}\over((1-T_2T)^{n-1}+T(T_2-T)^{n-1})((1-T_2T)^{m-1}+T(T_2-T)^{m-1})}\nonumber \\
&&=1+M_0(T_n T_m -T_n -T_m)=T_{m,n}.
\end{eqnarray}
We can calculate $\star$ product between states of the form $a^\dagger_k\cdots a^\dagger_l|n\rangle$ by differentiating eq.(\ref{eqn:MST}) with parameter $\beta$ and setting $\beta=0$.

\subsubsection{Identity and sliver state}

The matter identity state is given by
\begin{equation}
|I\rangle_M:=\mu_I\exp\left(-{1\over2}a^\dagger Ca^\dagger\right)|0\rangle,\ \ \ 
\mu_I\mu_3^M=\det{}^{d\over2}\left(1-M_0\right)=\det{}^{d\over2}\left({(1-T)^2\over1-T+T^2}\right),
\end{equation}
which is ``identity'' with respect to the $\star$ product in matter part:
\begin{equation}
{}_{3M}\langle I|V_3\rangle_{M123}={}_{34M}\langle V_2|I\rangle_{M4}|V_3\rangle_{M123}=|V_2\rangle_{M12}.
\end{equation}
This identity state corresponds to $n=1$ case in eqs.(\ref{eqn:MSQ})(\ref{eqn:TMN}) formally, and eqs.(\ref{eqn:MSQ}) (\ref{eqn:MST}) become
\begin{equation}
\label{eqn:IBETA}
|I_\beta\rangle=\mu_I\exp\left(\beta a^\dagger-{1\over2}a^\dagger Ca^\dagger\right)|0\rangle 
\end{equation}
and
\begin{eqnarray}
|I_\beta\star I_{\beta'}\rangle_M&=&e^{-{\cal C}_{1(\beta,\beta')}}|I_{\beta+\beta'}\rangle_M,\nonumber \\
{\cal C}_{1(\beta,\beta')}&=&{1\over2}\left(\beta{CM_-\over1-M_0}\beta^{'T}+\beta'{CM_+\over1-M_0}\beta^T\right)=\beta{CM_-\over1-M_0}\beta^{'T}=\beta'{CM_+\over1-M_0}\beta^T\nonumber \\
&=&{\cal C}_{1(\beta'C,\beta C)}.
\end{eqnarray}
respectively.\\

The matter sliver state $|\Xi\rangle_M$ is given by
\begin{equation}
|\Xi\rangle_M:=\mu_\infty \exp\left(-{1\over2}a^\dagger CTa^\dagger\right)|0\rangle,\ \ \ \mu_\infty=\det{}^{d\over2}\left(1-T^2\right)
\end{equation}
which is normalized as
\begin{equation}
|\Xi\star\Xi\rangle_M=|\Xi\rangle_M,\ \ \ \mu_3^M=\det{}^{d\over2}\left({1-T\over1-T+T^2}\right),\ \mu_I=\det{}^{d\over2}(1-T).
\end{equation}
This means $|\Xi\rangle_M$ is projection in matter part.

For consistency with $T_\infty=T$, suppose $(T-T_2)^n\rightarrow0, (n\rightarrow\infty)$, we can fix $\mu_2$ as
\begin{equation}
\mu_2=\det{}^{d\over2}(1-T_2T),
\end{equation}
and we get\footnote{
	$\mu_3=\mu_3^M$ if $T_2=0$ \cite{FO}.
}
\begin{equation}
\mu_n=\det{}^{d\over2}\left({1-T^2\over1+T\left({T_2-T\over1-T_2T}\right)^{n-1}}\right),\ \ \mu_1=\mu_I,\ \ ,|I\rangle_M=|1\rangle,\ |\Xi\rangle_M=|\infty\rangle.
\end{equation}
We have  the $\star$ product formula corresponding to $n,m=\infty$ in eq.(\ref{eqn:MST}) :
\begin{eqnarray}
&&|\Xi_\beta\star\Xi_{\beta'}\rangle=e^{-{\cal C}_{\infty(\beta,\beta')}}|\Xi_{\beta\rho_1+\beta'\rho_2}\rangle,\nonumber \\
&&{\cal C}_{\infty(\beta,\beta')}={1\over2}(\beta,\beta'){C\over(1-M_0)(1+T)}
\left(
\begin{array}{cc}
M_0(1-T)  & M_-  \\
M_+     & M_0(1-T)
\end{array}
\right)
\left(
\begin{array}{c}
\beta ^T    \\
\beta^{'T}     
\end{array}
\right)\nonumber \\
&&={1\over2}(\beta,\beta'){C\over1-T^2}
\left(
\begin{array}{cc}
T  & \rho_1-\rho_2T  \\
\rho_2-\rho_1T     & T
\end{array}
\right)
\left(
\begin{array}{c}
\beta ^T    \\
\beta^{'T}     
\end{array}
\right)={\cal C}_{\infty(\beta'C,\beta C)},\nonumber \\
&&\rho_1={M_-+TM_+\over(1+T)(1-M_0)}, \ \rho_2={M_++TM_-\over(1+T)(1-M_0)},\nonumber \\ 
&&{M_+\over1-M_0}={\rho_2-\rho_1T\over1-T},\ \ {M_-\over1-M_0}={\rho_1-\rho_2T\over1-T}.
\end{eqnarray}
Here $\rho_1,\rho_2$ are projection operators:
\begin{equation}
\rho_1^2=\rho_1,\ \rho_2^2=\rho_2,\ \rho_1+\rho_2=1,\ \rho_1\rho_2=\rho_2\rho_1=0,\ \rho_1C=C\rho_2.
\end{equation}

We also write down the $\star$ product between $n=1$ and $n=\infty$ state in eq.(\ref{eqn:MST}) :
\begin{eqnarray}
&&|I_\beta\star\Xi_{\beta'}\rangle=\exp\left(-{1\over2}\beta C{T\over1-T}\beta^T-\beta C{\rho_1-\rho_2T\over1-T}\beta^{'T}\right)|\Xi_{\beta\rho_1(1+T)+\beta'}\rangle,\nonumber \\
&&|\Xi_\beta\star I_{\beta'}\rangle=\exp\left(-{1\over2}\beta'C{T\over1-T}\beta^{'T}-\beta C{\rho_1-\rho_2T\over1-T}\beta^{'T}\right)|\Xi_{\beta+\beta'\rho_2(1+T)}\rangle.
\end{eqnarray}

\subsection{Ghost part}

Here we consider string field algebra in ghost part.
If we consider only ghost nonzero mode and use the Fock vacuum ${}_G\langle{\tilde +}|,\ |+\rangle_G$, we can get some similar formulas to those of matter part in previous subsection because Neumann coefficients $X^{rs}_{nm},\ n,m\geq1$ (\ref{eqn:VX3}) satisfy similar relations as (\ref{eqn:MMT}).  But because of ghost zero mode, the $\star$ product formulas are rather complicated than those in matter part. We introduce  {\it reduced} product and get some useful formulas and then we consider Witten's $\star$ product formula between ghost squeezed states in the Siegel gauge.\footnote{
	We call $|\Psi\rangle$ in the Siegel gauge if $b_0|\Psi\rangle=0$.
}

\subsubsection{Reduced product}

We define {\it reduced} product $\star^r$ by
\begin{equation}
\label{eqn:GSTRR}
|A\star^r B\rangle:= {}_2\langle A^r| {}_3\langle B^r|V^r_3\rangle_{123},\ \ \ \langle A^r|:=\langle V_2^r|A\rangle,
\end{equation}
where we restrict string fields $|A\rangle,|B\rangle$ such that they have no $b_0,c_0$ modes on the Fock vacuum $|+\rangle$.
Here we introduced reduced reflector $\langle V_2^r|$ and reduced 3-string vertex $|V_3^r\rangle$ which contain no $b_0,c_0$ modes on the vacuum ${}_G\langle{\tilde +}|,\ |+\rangle_G$, i.e. they are related with usual reflector (\ref{eqn:RFB}) and 3-string vertex (\ref{eqn:3SV}) by
\begin{equation}
{}_{12}\langle V_2| = {}_{12}\langle V_2^r|(c_0^{(1)}+c_0^{(2)}),\ \ 
|V_3\rangle_{123}=\exp\left(-\sum_{r,s=1}^3c^{\dagger(r)}X^r{}^s_0b^{(s)}_0\right)|V_3^r\rangle_{123}.
\end{equation}
This $\star^r$ product satisfies associativity : $|(A \star^r B)\star^r C\rangle=|A \star^r (B\star^r C)\rangle$. In fact we have obtained the following formula by straightforward calculation :
\begin{eqnarray}
&&{}_{26}\langle V_2^r|V_3^r\rangle_{456}|V_3^r\rangle_{312}= {}_{26}\langle V_2^r|V_3^r\rangle_{536}|V_3^r\rangle_{142}={}_{26}\langle V_2^r|V_3^r\rangle_{146}|V_3^r\rangle_{532}\nonumber \\
&&= (\tm_3^r)^2 \det (1-\M0^2) e^{E_{1345}}|+\rangle_{G1345},\nonumber \\
&&E_{1345}=\sum_{s,t=1,3,4,5}c^{(s)\dagger}E^{st}b^{(t)\dagger},\nonumber \\
&&E^{11}=E^{33}=E^{44}=E^{55}={C\M0 \over1+\M0 },\ \ 
E^{13}=E^{35}=E^{41}=E^{54}={C\Mm^2 \over1-\M0^2 },\nonumber \\
&&E^{14}=E^{31}=E^{45}=E^{53}={C\Mp^2 \over1-\M0^2},\ \ 
E^{15}=E^{34}=E^{43}=E^{51}={C\Mp \Mm \over1-\M0^2}.
\end{eqnarray}

We define ghost squeezed state $|n_{\xi,\eta}\rangle$ with Grassmann odd parameters $\xi,\eta$ which corresponds to $|n_\beta\rangle$ (\ref{eqn:MSQ}) in matter part :
\begin{equation}
\label{eqn:NXH}
|n_{\xi,\eta}\rangle:=e^{\xi b^\dagger+\eta c^\dagger}|n\rangle_G=\tm_n \exp\left(\xi b^\dagger+\eta c^\dagger+c^\dagger C\T_n b^\dagger\right)|+\rangle_G, 
\end{equation}
where $|n\rangle_G$ is defined by the state which is obtained by taking the $\star^r$ product $n-1$ times with a particular ghost squeezed state $|2\rangle_G$ :
\begin{equation}
\label{eqn:GNWED}
|n\rangle_G=\left(|2\rangle_G\right)_{\star^r}^{n-1},\ \ \ |2\rangle_G=\exp\left(c^\dagger C\T_2 b^\dagger\right)|+\rangle_G.
\end{equation}
We take a matrix $\T_2$ which satisfies
\begin{equation}
C\T_2=\T_2C,\ \ [\M0 ,\T_2]=0,\ \ \T_2\not=1,
\end{equation}
for simplicity, and then we have obtained formulas for $\T_n,\tm_n$,
\begin{eqnarray}
\label{eqn:TNTIL}
\T_n &=&{\T (1-\T_2 \T)^{n-1}+(\T_2-\T )^{n-1}\over(1-\T_2 \T)^{n-1}+\T (\T_2-\T )^{n-1}}, \nonumber\\
\tm_n &=& \tm_2 \left(\tm_2 \tm_3^r \det\left({1-\T \over1-\T+\T^2}\right)\right)^{n-2}\det\left({(1-\T_2 \T )^{n-1}+\T(\T_2- \T )^{n-1}\over1-\T^2 }\right),\nonumber \\
\M0 &=&{\T\over1-\T+\T^2}.
\end{eqnarray}
by solving the same recurrence equation as that in matter part (\ref{eqn:TMN}).
Here $\tm_3^r$ is normalization factor of reduced 3-string vertex  $|V_3^r\rangle$ and $\T$ is expressed with $\M0$ as a solution of the quadratic equation
$\M0 \T^2-(1+\M0 )\T +\M0=0$.

For these ghost squeezed states, we have the $\star^r$ product formula:
\begin{equation}
\label{eqn:GNMSTRR}
|n_{\xi,\eta}\star^r m_{\xi',\eta'}\rangle=\exp\left(-{\cal C}_{n_{\xi,\eta},m_{\xi',\eta'}}\right)|(n+m-1)_{\xi{\tilde \rho}_{1(n,m)}+\xi'{\tilde \rho}_{2(n,m)},\eta{\tilde \rho}_{1(n,m)}^T+\eta'{\tilde \rho}_{2(n,m)}^T}\rangle,
\end{equation}
where
\begin{eqnarray}
%&&\langle n_{\xi,\eta}^r|=\langle n^r|e^{\xi Cb-\eta Cc}=\tm_n{}_G\langle{\tilde +}|\exp\left(-c\T_n C b+\xi Cb-\eta Cc\right),\nonumber\\
&&{\cal C}_{n_{\xi,\eta},m_{\xi',\eta'}}=(\xi,\xi'){C\over\T_{n,m}}
\left(
\begin{array}{cc}
\M0 (1-\T_m)  & \Mm  \\
\Mp     & \M0 (1-\T_n)
\end{array}
\right)
\left(
\begin{array}{c}
\eta^T     \\
\eta^{'T}     
\end{array}
\right)={\cal C}_{m_{\xi'C,\eta'C},n_{\xi C,\eta C}},\nonumber \\
&&{\tilde \rho}_{1(n,m)}={\Mm +\Mp \T_m\over\T_{n,m}},\ \ {\tilde \rho}_{2(n,m)}={\Mp +\Mm \T_n\over\T_{n,m}},\ \ C{\tilde \rho}_{1(n,m)}={\tilde \rho}_{2(m,n)}C, \nonumber \\
&&\T_{n,m}={(1+\T )(1-\T )^2\over1-\T +\T^2} {(1-\T_2 \T)^{n+m-2}+\T (\T_2-\T )^{n+m-2}\over((1-\T_2 \T )^{n-1}+\T (\T_2- \T )^{n-1})((1-\T_2 \T )^{m-1}+\T (\T_2 -\T )^{m-1})}\nonumber \\
&&=1+\M0 (\T_n \T_m-\T_n-\T_m)=\T_{m,n}.
\end{eqnarray}
These formulas are similar to those in matter part (\ref{eqn:MST})(\ref{eqn:MSTP}). We can obtain $\star^r$ product between the states of the form $b_k^\dagger\cdots c^\dagger_l|n\rangle_G$ by differentiating eq.(\ref{eqn:GNMSTRR}) with respect to parameters $\xi,\eta$ and setting them zero.

\subsubsection{Identity and sliver like state}

We consider a particular state $|I^r\rangle$ which was excluded as a candidate for $|2\rangle_G$ :
\begin{eqnarray}
|I^r\rangle_G:=\tm_I^r\exp\left(c^\dagger C b^\dagger\right)|+\rangle_G,\ \ \ 
\tm_I^r \tm_3^r=\det{}^{-1}\left(1-\M0 \right)=\det\left({1-\T+\T^2\over(1-\T )^2}\right).
\end{eqnarray}
This is identity-like state of ghost part,i.e., identity with respect to the $\star^r$ product which was introduced in eq.(\ref{eqn:GSTRR}) because it satisfies the following equation\footnote{
Here the ket $|V_2^r\rangle$ is reduced reflector which was defined in eq.(\ref{eqn:IGP}). Note that ${}_{12}\langle V_2^r|V_2^r\rangle_{23}=1_{31}$ if we consider string fields which have no $b_0,c_0$ modes on the vacuum ${}_G\langle{\tilde +}|,\ |+\rangle_G$.
}
\begin{equation}
{}_3\langle I^r|V_3^r\rangle_{123}= {}_{34}\langle V_2^r|I^r\rangle_4 |V_3^r\rangle_{123}= |V_2^r\rangle_{12},
\end{equation}
although it is {\it different} from the identity state $|I\rangle$ (\ref{eqn:IKE}) in ghost part as :
\begin{equation}
|I\rangle={1\over4i}b^+\left({\pi\over2}\right)b^-\left({\pi\over2}\right)|I\rangle_M|I^r\rangle_G=[b^\dagger]_{\cal O} (b_0+2 [b^\dagger]_{\cal E}) |I\rangle_M|I^r\rangle_G,
\end{equation}
where we used the notation of (\ref{eqn:OESUM}).
Corresponding to  (\ref{eqn:IBETA}) in matter part, we consider the state of the form
\begin{equation}
|I_{\xi,\eta}\rangle=\tm_I^r\exp\left(\xi b^\dagger+\eta c^\dagger+c^\dagger Cb^\dagger\right)|+\rangle_G,
\end{equation}
then we have the $\star^r$ product formula between them :
\begin{eqnarray}
|I_{\xi,\eta}\star^rI_{\xi',\eta'}\rangle&=&e^{-{\cal C}_{1(\xi,\eta;\xi',\eta')}}|I_{\xi+\xi',\eta+\eta'}\rangle,\nonumber \\
{\cal C}_{1(\xi,\eta;\xi',\eta')}&=&\xi{C\Mm \over1-\M0 }\eta^{'T }+\xi'{C\Mp \over1-\M0 }\eta^T={\cal C}_{1(\xi'C,\eta'C;\xi C,\eta C)}.
\end{eqnarray}
This is $n=m=1$ case of eq.(\ref{eqn:GNMSTRR}) formally.\\

Next we define the sliver-like state in ghost part :
\begin{equation}
|\Xi^r\rangle_G:=\tm_\infty \exp\left(c^\dagger C\T b^\dagger\right)|+\rangle_G,\ \ \tm_\infty=\det{}^{-1}\left(1-\T^2\right).
\end{equation}
This state is analogy of projection with respect to the  $\star^r$ product :
\begin{equation}
|\Xi^r\star^r\Xi^r\rangle_G=|\Xi^r\rangle_G,\ \ 
\tm_3^r=\det{}^{-1}\left({1-\T \over1-\T +\T^2}\right),\ \ 
\tm_I^r=\det{}^{-1}(1-\T ).
\end{equation}
For formal consistency with the case $n=\infty$ in eqs.(\ref{eqn:NXH})(\ref{eqn:TNTIL}), suppose $(\T -\T_2)^n\rightarrow0\  (n\rightarrow\infty)$, we fix $\tm_2$ as
\begin{equation}
\tm_2=\det{}^{-1}(1-\T_2 \T ),
\end{equation}
and then we get\footnote{
	$\tm_3=\tm_3^r$ if $\T_2=0$.
}
\begin{equation}
\tm_n=\det{}^{-1}\left({1-\T^2\over1+\T \left({\T_2-\T \over1-\T_2\T }\right)^{n-1}}\right),\ \ \tm_1=\tm_I^r,\ \ ,|I^r\rangle_G=|1\rangle_G,\ |\Xi^r\rangle=|\infty\rangle_G.
\end{equation}
We have obtained the $\star^r$ product formula for $n,m=\infty$ case in eq.(\ref{eqn:GNMSTRR}) :
\begin{eqnarray}
&&|\Xi_{\xi,\eta}\star^r\Xi_{\xi',\eta'}\rangle=e^{-{\cal C}_{\infty(\xi,\eta;\xi',\eta')}}|\Xi_{\xi{\tilde \rho}_1+\xi'{\tilde \rho}_2,\eta{\tilde \rho}_1^T+\eta'{\tilde \rho}_2^T}\rangle,\nonumber \\
&&{\cal C}_{\infty(\xi,\eta;\xi',\eta')}=(\xi,\xi'){C\over(1-\M0 )(1+\T )}
\left(
\begin{array}{cc}
\M0 (1-\T )  & \Mm  \\
\Mp     & \M0 (1-\T )
\end{array}
\right)
\left(
\begin{array}{c}
\eta^T    \\
\eta^{'T}     
\end{array}
\right)\nonumber \\
&&=(\xi,\xi'){C\over1-\T^2}
\left(
\begin{array}{cc}
\T  & {\tilde \rho}_1-{\tilde \rho}_2\T  \\
{\tilde \rho}_2-{\tilde \rho}_1\T     & \T
\end{array}
\right)
\left(
\begin{array}{c}
\eta^T    \\
\eta^{'T}     
\end{array}
\right)
={\cal C}_{\infty(\xi'C,\eta'C;\xi C,\eta C)},\nonumber \\
&&{\tilde \rho}_1={\Mm +\T \Mp \over(1+\T )(1-\M0 )},\ {\tilde \rho}_2={\Mp +\T \Mm \over(1+\T )(1-\M0 )},\nonumber \\
&&{\Mp\over1-\M0 }={{\tilde \rho}_2-{\tilde \rho}_1\T \over1-\T },\ {\Mm\over1-\M0 }={{\tilde \rho}_1-{\tilde \rho}_2\T \over1-\T }.\nonumber \\
\end{eqnarray}
In this notation, ${\tilde \rho}_1,{\tilde \rho}_2$ are projection operators :
\begin{equation}
{\tilde \rho}_1^2={\tilde \rho}_1,\ {\tilde \rho}_2^2={\tilde \rho}_2,\ {\tilde \rho}_1+{\tilde \rho}_2=1,\ {\tilde \rho}_1{\tilde \rho}_2={\tilde \rho}_2 {\tilde \rho}_1=0,\ C{\tilde \rho}_1={\tilde \rho}_2 C.
\end{equation}
We also write down the $\star^r$ product between the identity and sliver like states :
\begin{eqnarray}
&&|I_{\xi,\eta}\star^r\Xi_{\xi',\eta'}\rangle\nonumber \\
&&=\exp\left(-\xi C{\T\over1-\T }\eta^T-\xi C{{\tilde \rho}_1-{\tilde \rho}_2\T \over1-\T }\eta^{'T}-\xi'C{{\tilde \rho}_2-{\tilde \rho}_1\T \over1-\T }\eta^T\right)|\Xi_{\xi{\tilde \rho}_1(1+\T )+\xi',\eta({\tilde \rho}_1(1+\T ))^T+\eta'}\rangle,\nonumber \\
&&|\Xi_{\xi,\eta}\star^rI_{\xi',\eta'}\rangle\nonumber \\
&&=\exp\left(-\xi'C{\T \over1-\T }\eta^{'T}-\xi C{{\tilde \rho}_1-{\tilde \rho}_2\T \over1-\T }\eta^{'T}-\xi'C{{\tilde \rho}_2-{\tilde \rho}_1\T \over1-\T }\eta^T\right)|\Xi_{\xi+\xi'{\tilde \rho}_2(1+\T ),\eta+\eta'({\tilde \rho}_2(1+\T ))^T}\rangle.\nonumber \\
\end{eqnarray}

\subsubsection{The $\star$ product in ghost part}

Here we consider Witten's $\star$ product defined by eq.(\ref{eqn:STR}) in the Siegel gauge.
By using the reduced product $\star^r$ (\ref{eqn:GSTRR}), the $\star$ product (\ref{eqn:STR}) in the Siegel gauge can be written rather simply, i.e. for $|\Phi\rangle=b_0|\phi\rangle,|\Psi\rangle=b_0|\psi\rangle$, where $|\phi\rangle,|\psi\rangle$ have no $b_0,c_0$ modes, their $\star$ product is\footnote{
	We supposed that $|\phi\rangle,|\psi\rangle$ are Grassmann even as $|n\rangle_G$ in this formula.
}
\begin{eqnarray}
\label{eqn:STRSTRR}
|\Phi\star\Psi\rangle&=&|\phi\star^r\psi\rangle+b_0\left({}_2\langle\phi^r|{}_3\langle\psi^r|\sum_{s=1}^3c^{(s)\dagger}X^s{}^1_0|V_3^r\rangle_{123}\right)\nonumber \\
&=&(1+b_0c^\dagger X^1{}^1_0)|\phi\star^r\psi\rangle+b_0\sum_{s=2,3} {}_2\langle\phi^r |{}_3\langle\psi^r|c^{(s)\dagger}X^s{}^1_0 |V^r_3\rangle_{123} .
\end{eqnarray}
From this formula (\ref{eqn:STRSTRR}) and eq.(\ref{eqn:GNMSTRR}), we have obtained the $\star$ product between ghost squeezed states (\ref{eqn:NXH}) in the Siegel gauge :
\begin{eqnarray}
&&|(b_0 n_{\xi,\eta})\star(b_0 m_{\xi',\eta'})\rangle\nonumber \\
&=&\left(1+b_0\left(c^\dagger X^1{}^1_0+\left(\xi C+{\partial\over\partial\eta}\T_n\right)X^2{}^1_0
+\left(\xi'C+{\partial\over\partial\eta'}\T_m\right)X^3{}^1_0\right)\right)|n_{\xi,\eta}\star^r m_{\xi',\eta'}\rangle\nonumber \\
&=&\left(1+b_0c^\dagger{1-\T_n \T_m\over\T_{n,m}}X^1{}^1_0-b_0(\xi{\tilde \rho}_{1(n,m)}+\xi'{\tilde \rho}_{2(n,m)}){1\over1-\M0 }X^1{}^1_0\right)|n_{\xi,\eta}\star^r m_{\xi',\eta'}\rangle,
\end{eqnarray}
where
\begin{equation}
{1-\T_n \T_m\over\T_{n,m}}={1-\T \over1-\M0 } \cdot {(1-\T_2\T)^{n+m-2}-(\T_2- \T )^{n+m-2} \over (1-\T_2 \T)^{n+m-2}+\T (\T_2- \T )^{n+m-2}}.
\end{equation}
In particular, for the identity and sliver like states we get some rather simple formulas:
\begin{eqnarray}
&&|b_0I^r\star b_0I^r\rangle=|I^r\rangle,\nonumber \\
&&|b_0\Xi^r\star b_0\Xi^r\rangle=|b_0I^r\star b_0\Xi^r\rangle=|b_0\Xi^r\star b_0I^r\rangle=\left(1+b_0c^\dagger{1-\T \over1-\M0 }X^1{}^1_0\right)|\Xi^r\rangle.
\end{eqnarray}
and with parameters $\xi,\eta$,
\begin{eqnarray}
\label{eqn:ISXH}
&&|b_0I_{\xi,\eta}\star b_0I_{\xi',\eta'}\rangle%\nonumber \\
%&&=\left(1+b_0\left(\left(\xi C+\xi'C{\Mp \over1-\M0}\right)X^2{}^1_0+\left(\xi'C+\xi C{\Mm \over1-\M0}\right)X^3{}^1_0\right)\right)|I_{\xi,\eta}\star^r I_{\xi',\eta'}\rangle\nonumber \\
%&&=\left(1+b_0\left(\left(\xi C+\xi'C{{\tilde \rho}_2-{\tilde \rho}_1\T  \over1-\T }\right)X^2{}^1_0+\left(\xi'C+\xi C{{\tilde \rho}_1-{\tilde \rho}_2\T  \over1-\T }\right)X^3{}^1_0\right)\right)|I_{\xi,\eta}\star^r I_{\xi',\eta'}\rangle,\nonumber \\
=\left(1-b_0 (\xi+\xi'){1\over1-\M0 }X^1{}^1_0\right)e^{-{\cal C}_{1(\xi,\eta;\xi',\eta')}}|I_{\xi+\xi',\eta+\eta'}\rangle,\nonumber \\
&&|b_0\Xi_{\xi,\eta}\star b_0\Xi_{\xi',\eta'}\rangle\nonumber \\
%&&=\biggl(1+b_0c^\dagger(X^1{}^1_0+{\tilde \rho}_1\T X^2{}^1_0+{\tilde \rho}_2\T X^3{}^1_0)+b_0\biggl(\biggl(\xi C{1\over1-\T^2}+\xi'C\T {\Mp \over(1+\T )(1-\M0 )}\biggr)X^2{}^1_0
%\nonumber \\
%&&+\biggl(\xi'C{1\over1-\T^2}+\xi C\T {\Mm \over(1+\T )(1-\M0 )}\biggr)X^3{}^1_0\biggr)\biggr)|\Xi_{\xi,\eta}\star^r \Xi_{\xi',\eta'}\rangle\nonumber \\
%&&=\biggl(1+b_0c^\dagger(X^1{}^1_0+{\tilde \rho}_1\T X^2{}^1_0+{\tilde \rho}_2\T X^3{}^1_0)\nonumber \\
%&&+b_0\biggl(\biggl(\xi C{1\over1-\T^2}+\xi'C\T {{\tilde \rho}_2-{\tilde \rho}_1\T  \over1-\T^2 }\biggr)X^2{}^1_0
%+\biggl(\xi'C{1\over1-\T^2}+\xi C\T {{\tilde \rho}_1-{\tilde \rho}_2\T  \over1-\T^2 }\biggr)X^3{}^1_0\biggr)\biggr)|\Xi_{\xi,\eta}\star^r \Xi_{\xi',\eta'}\rangle,\nonumber \\
&&=\left(1+b_0c^\dagger{1-\T \over1-\M0 }X^1{}^1_0 -b_0(\xi{\tilde \rho}_1+\xi'{\tilde \rho}_2){1\over1-\M0 }X^1{}^1_0 \right)e^{-{\cal C}_{\infty(\xi,\eta;\xi',\eta)}}|\Xi_{\xi{\tilde \rho}_1+\xi'{\tilde \rho}_2,\eta{\tilde \rho}_1^T+\eta'{\tilde \rho}_2^T}\rangle,\nonumber \\
&&|b_0I_{\xi,\eta}\star b_0\Xi_{\xi',\eta'}\rangle%=\biggr(1+b_0c^\dagger({\tilde \rho}_2(1+\T) \T -1 )X^3{}^1_0\nonumber\\
%&&+b_0\biggl(\biggl(\xi C{1\over1-\T }+\xi'C{{\tilde \rho}_2-{\tilde \rho}_1\T \over1-\T }\biggr)X^2{}^1_0
%+\biggl(\xi C\T {{\tilde \rho}_1-{\tilde \rho}_2\T \over1-\T } + \xi'C \biggr)X^3{}^1_0\biggr) \biggr)|I_{\xi,\eta}\star^r \Xi_{\xi',\eta'}\rangle\nonumber\\
=\left(1+b_0c^\dagger{1-\T \over1-\M0 }X^1{}^1_0-b_0(\xi{\tilde \rho}_1(1+\T )+\xi'){1\over1-\M0 }X^1{}^1_0\right)|I_{\xi,\eta}\star^r \Xi_{\xi',\eta'}\rangle,\nonumber \\
&&|b_0\Xi_{\xi,\eta}\star b_0I_{\xi',\eta'}\rangle 
%=\biggr(1+b_0c^\dagger({\tilde \rho}_1(1+\T) \T -1 )X^2{}^1_0\nonumber\\
%&&+b_0\biggl(\biggl(\xi C+\xi'C\T {{\tilde \rho}_2-{\tilde \rho}_1\T \over1-\T }\biggr)X^2{}^1_0
%+\biggl(\xi C{{\tilde \rho}_1-{\tilde \rho}_2\T \over1-\T }  + \xi'C{1\over1-\T } \biggr)X^3{}^1_0\biggr) \biggr)|\Xi_{\xi,\eta}\star^r I_{\xi',\eta'}\rangle\nonumber \\
=\left(1+b_0c^\dagger{1-\T \over1-\M0 }X^1{}^1_0-b_0(\xi+\xi'{\tilde \rho}_2 (1+\T )){1\over1-\M0 }X^1{}^1_0\right)|\Xi_{\xi,\eta}\star^r I_{\xi',\eta'}\rangle.\nonumber \\
\end{eqnarray}

\section{Solutions of vacuum string field theory \label{sec:VSFT}}

We consider equation of motion of VSFT \cite{VSFT} for an application of techniques which we have developed in previous sections.

The equation of motion of VSFT is given by
\begin{equation}
\label{eqn:EOMVSFT}
Q|\Psi\rangle+|\Psi\star\Psi\rangle=0,\ \ ,
\end{equation}
where  $Q$ is given by a linear combination of $c$ ghost modes :
\begin{equation}
\label{eqn:VSFTQ}
Q=c_0+\sum_{n=1}^\infty f_n(c_n+(-1)^n c^\dagger_n)=c_0+f\cdot (c+Cc^\dagger).
\end{equation}
Here $f_n$s are particular coefficients which we choose appropriately.
To solve eq.(\ref{eqn:EOMVSFT}), we set the ansatz for solutions in the Siegel gauge as :
\begin{equation}
\label{eqn:ANSATZ}
|\Psi\rangle=b_0|P\rangle_M\left(\sum_{n=1}^\infty g_n |n \rangle_G\right),\ \ |P\star P\rangle_M=|P\rangle_M.
\end{equation}
This means that it is factorized into matter and ghost part, matter part is some projector and ghost part is some linear combination of ghost squeezed states of the form (\ref{eqn:NXH}) with appropriate coefficients $g_n$ for simplicity.
In matter part, we can choose $|P\rangle_M$ as matter identity $|I\rangle_M$ or matter sliver state $|\Xi\rangle_M$ for example.\footnote{
	If we use half-string formulation \cite{HALF}\cite{RSZ}\cite{KOM}, we can choose other projector in matter part.
} To solve eq.(\ref{eqn:EOMVSFT}), we substitute (\ref{eqn:ANSATZ}) and find coefficients $g_n$ of $|\Psi\rangle$ and  $f_n$ of $Q$ in the same time.

Noting
\begin{equation}
Q|n\rangle_G=c^\dagger C(1-\T_n)\cdot f^T |n\rangle_G,
\end{equation}
we have obtained the following solutions.
\begin{enumerate}
\item identity-like solution
\begin{equation}
Q=c_0,\ \ |\Psi\rangle=-b_0|P\rangle_M |I^r\rangle_G.
\end{equation}

\item sliver-like solution 
\begin{equation}
Q=c_0-(c+c^\dagger){1\over1-\M0 }X^1{}^1_0,\ \ |\Psi\rangle=-b_0|P\rangle_M |\Xi^r\rangle_G.
\end{equation}
This was constructed in \cite{HK}.\footnote{
	This formula is simpler than that in \cite{HK}.
}

\item another solution
\begin{equation}
Q=c_0-(c+c^\dagger){1\over1-\M0 }X^1{}^1_0,\ \ |\Psi\rangle=-b_0|P\rangle_M(|I^r\rangle_G-|\Xi^r\rangle_G).
\end{equation}
This is a solution for the $Q$ which is the same as the above one.
\end{enumerate}

The chosen $Q$s for these solutions consist of even modes of $c$ ghost because
\begin{equation}
(X^1{}^1_0)_{2n+1}=0,\ \ \ (\M0)_{2n,2m+1}=(\M0)_{2m+1,2n}=0
\end{equation}
which are obtained from eqs.(\ref{eqn:MMT})(\ref{eqn:VXTT}).
Although these $Q$ do not vanish on the identity state $\langle I|$ (\ref{eqn:IBR}) :
\begin{equation}
\langle I|c_0=\langle I|{(-1)^n\over2}(c_{2n}+c_{2n}^\dagger)\not=0
\end{equation}
we have obtained
\begin{equation}
\label{eqn:C0IV}
{}_3\langle I|c_0^{(3)}|V_3\rangle_{123}=0
\end{equation}
by straightforward calculation. 
This means that we could use $Q$ which consists of an arbitrary linear combination of $c_0,(c_{2n}+c^\dagger_{2n})$ for the definition of VSFT in our oscillator formulation although it was afraid to be used if $\langle I|Q\not=0$ in \cite{VSFT}.\\

Note that the equation (\ref{eqn:C0IV}) also means
\begin{equation}
\label{eqn:COIA}
|(c_0I)\star A\rangle=0,\ \ \forall |A\rangle.
\end{equation}
If we take identity state $|I\rangle$ (\ref{eqn:IKE}) as $|A\rangle$, this becomes $|(c_0I)\star I\rangle=0$. If identity state $|I\rangle$ were identity with respect to the $\star$ product, this might show $|c_0I\rangle=0$, but it is inconsistent with the fact $c_0|I\rangle\not=0$. We can solve this mystery which was referred in \cite{RZ} in our oscillator formulation.
Because identity state $|I\rangle$ is not identity  as was shown in \S \ref{sec:WITI}, we have \\ $|(c_0I)\star I\rangle\not=c_0|I\rangle$ consistently. Eq.(\ref{eqn:COIA}) also follows from the fact
\begin{equation}
 (c_0^{(1)}+c_0^{(2)}+c_0^{(3)})|V_3\rangle=0,\ \ c_0|I\star A\rangle=|I\star(c_0A)\rangle,\ \ \forall A
\end{equation}
which we can check from eqs.(\ref{eqn:3SV})(\ref{eqn:IV2I}).

\section{Discussion \label{sec:DIS}}

In this paper we have constructed some squeezed states $|n_{\xi,\eta}\rangle$ with parameters $\xi,\eta$ which correspond to wedge states in matter part. They satisfy rather simple algebra with respect to the reduced $\star^r$ product which we introduced for convenience. We have also obtained the $\star$ product formula between them in the Siegel gauge. 

For ghost squeezed states, the $\star^r$ product formulas are very similar to those in matter part, but the $\star$ product is a little complicated by ghost zero mode.
In these calculations, the algebras which Neumann coefficients satisfy are essential. We have shown them first with some review for self-contained.

By applying our formulations to VSFT, we have obtained some solutions of equation of motion which are expressed by the ghost identity and sliver like states. To get some physical implications, we should examine these solutions as was done for a particular solution in \cite{HK}. It is a future problem to discuss physical spectrum around these solutions, potential hight and so on.
We hope that our techniques will be helpful to get solutions of original Witten's string field theory and analyze them.

Although we have used oscillator representation consistently and performed algebraic calculations only, it is also important to interpret them in the CFT language to obtain geometrical meaning of our treatment in ghost part.

We noted that the identity state $|I\rangle$ is not identity with respect to the $\star$ product by straightforward calculation. 
We can also show that associativity of the $\star$ product is broken generally by calculating  ${}_{26}\langle V_2|V_3\rangle_{456}|V_3\rangle_{312}$  straightforwardly although the reduced $\star^r$ product satisfies associativity. This comes from Neumann coefficients $X^r{}^s_0$ which are related to ghost zero mode. This problem which is inconsistent with naive consideration about connection condition was known as associativity anomaly.\cite{STR}  It is subtle but important problem to discuss Witten type string field theory rigorously and consistently.\footnote{
It was not necessary to consider associativity of the $\star$ product
in this paper because we discussed only equation of motion. 
} It is a future problem to investigate this issue more seriously.

\section*{Acknowledgements}
We would like to thank H.~Hata, T.~Kawano, T.~Kugo,  Y.~Matsuo, K.~Okuyama, T.~Takahashi and S.~Teraguchi for valuable discussions and comments.
We wish to thank K.~Ohmori for encouragements.

We thank the organizers of Summer Institute 2001 at Yamanashi
for its stimulating atmosphere where this work has begun.

\appendix

\section{Conventions}

Here we list up our conventions.

Mode expansion in matter part ($\mu=1,\cdots,d$):
\begin{eqnarray}
X^\mu(\sigma)&=&x_0^\mu+\sqrt2 \sum_{n=1}^\infty x_n^\mu\cos n\sigma=x_0^\mu+i\sqrt{2\alpha'}\sum_{n=1}^\infty{1\over n}(\alpha_n^\mu-\alpha_{-n}^\mu)\cos n\sigma,\nonumber \\
P_\mu(\sigma)&=&{1\over\pi}\left(p_{0\mu}+\sum_{n=1}^\infty p_{n\mu}\cos n\sigma\right)={1\over\pi}\left(p_{0\mu}+{1\over\sqrt{2\alpha'}}\sum_{n=1}^\infty\eta_{\mu\nu}(\alpha_n^\nu+\alpha_{-n}^\nu)\cos n\sigma\right).\nonumber \\
\end{eqnarray}
Matter nonzero mode ($n\geq1$):
\begin{eqnarray}
&&\alpha_n^\mu=\sqrt n a_n^\mu,\ \alpha_{-n}^\mu=\sqrt{n}a_n^{\dagger\mu},\ \ [\alpha_n^\mu,\alpha_m^\nu]=n\delta_{n+m,0}\eta^{\mu\nu},\ [a_n^\mu,a_m^{\dagger\nu}]=\delta_{n,m}\eta^{\mu\nu},\nonumber \\
&&x_n^\mu={i\over\sqrt{2n}}\sqrt{2\alpha'}(a_n^\mu-a_n^{\dagger\mu}),\ p_{n\mu}=\sqrt{n\over2}{\eta_{\mu\nu}\over\sqrt{2\alpha'}}(a_n^\nu+a_n^{\dagger\nu}),\ \ [x_n^\mu,p_{m\nu}]=i\delta_{n,m}\delta^\mu_\nu.
\end{eqnarray}
Matter zero mode:
\begin{equation}
x_0^\mu={i\over2}\sqrt{2\alpha'}(a_0^\mu-a_0^{\dagger\mu}),\ p_{0\mu}={1\over\sqrt{2\alpha'}}\eta_{\mu\nu}(a_0^\nu+a_0^{\dagger\nu}),\ \ [x_0^\mu,p_{0\nu}]=i\delta^\mu_\nu,\ [a_0^\mu,a_0^{\dagger\nu}]=\eta^{\mu\nu}.
\end{equation}
Mode expansion of $bc$ ghost :
\begin{eqnarray}
c^{\pm}(\sigma)=\sum_{n=-\infty}^\infty c_n e^{\mp in\sigma},\ \ b^{\pm}(\sigma)=\sum_{n=-\infty}^\infty b_n e^{\mp in\sigma},\ \ \ \{c_n,b_m\}=\delta_{n+m,0}.
\end{eqnarray}

\section{Some formulas}

Here we collect some useful formulas which we often use in calculations.

\subsection{Gaussian integral}

We can prove useful formulas for oscillators by using coherent state and performing Gaussian integral.

For bosonic oscillator $a_n,a_n^\dagger$ and Fock vacuum $|0\rangle$ :
\begin{equation}
[a_m,a_n^\dagger]=\delta_{mn},\ a_n|0\rangle=0,
\end{equation}
we get a formula
\begin{eqnarray}
&&\exp\left({1\over2}aMa+\lambda a\right)\exp\left({1\over2}a^\dagger Na^\dagger+\mu a^\dagger\right)|0\rangle\nonumber\\
&&={1\over\sqrt{{\rm det}(1-MN)}}\exp\left({1\over2}\lambda N(1-MN)^{-1}\lambda+{1\over2}\mu M(1-NM)^{-1}\mu+\lambda(1-NM)^{-1}\mu\right)\nonumber \\
&&\cdot \exp\left((\lambda N+\mu)(1-MN)^{-1}a^\dagger+{1\over2}a^\dagger N(1-MN)^{-1}a^\dagger\right)|0\rangle,
\end{eqnarray}
where $M,N$ are symmetric matrices.

For fermionic oscillator\footnote{
	We often use the notation :  $b_{-n}=b_n^\dagger,c_{-n}=c_n^\dagger,\ n\geq1$.
} $b_n,c_n,b_n^\dagger,c_n^\dagger$ and Fock vacuum $|+\rangle_G$ :
\begin{equation}
\{b_m,c_n^\dagger\}=\delta_{mn},\ \{c_m,b_n^\dagger\}=\delta_{mn},\ c_n|+\rangle_G=b_n|+\rangle_G=0,
\end{equation}
we get\footnote{
	Note that this formula contains no $b_0,c_0$.
}
\begin{eqnarray}
&&\exp\left(bMc+b\lambda+\mu c\right)\exp\left(b^\dagger Nc^\dagger+b^\dagger\xi+\eta c^\dagger\right)|+\rangle_G\nonumber \\
&&={\rm det}(1+MN)\exp\left((\mu-\eta M)(1+NM)^{-1}\xi-(\mu N+\eta)(1+MN)^{-1}\lambda\right)\nonumber \\
&&\cdot \exp\left(b^\dagger(1+NM)^{-1}(-N\lambda+\xi)+(\mu N+\eta)(1+MN)^{-1}c^\dagger+b^\dagger N(1+MN)^{-1}c^\dagger\right)|+\rangle_G,\nonumber \\
\end{eqnarray}
where $M,N$ are Grassmann even and $\lambda,\mu,\xi,\eta$ are odd.\\

\subsection{Coherent states for $bc$-ghost}

In our concrete calculations which contain both $bc$ ghost nonzero and zero modes, it is useful to insert completeness formula for coherent states.

Coherent states of nonzero mode:
\begin{eqnarray}
&&|\xi\rangle:=e^{-\xi^c b^\dagger-\xi^b c^\dagger}|+\rangle_G,\ \  \langle\xi|:= {}_G\langle{\tilde +}|e^{-b{\bar \xi}^c-c{\bar \xi}^b},\nonumber \\
&&b_n|\xi\rangle=\xi^b_n|\xi\rangle,\ \ \langle\xi|b^\dagger_n=\langle\xi|{\bar \xi}^b_n,\ \ \ c_n|\xi\rangle=\xi^c_n|\xi\rangle,\ \ \langle\xi|c^\dagger_n=\langle\xi|{\bar \xi}^c_n.
\end{eqnarray}
Coherent state of zero mode:
\begin{eqnarray}
|\xi_0\rangle:=e^{-\xi_0b_0}|+\rangle_G,&& \langle\xi_0|:= {}_G\langle{\tilde +}|e^{-c_0{\bar \xi}_0},\nonumber \\
c_0|\xi_0\rangle=\xi_0|\xi_0\rangle, && \langle\xi_0|b_0=\langle\xi_0|{\bar \xi}_0.
\end{eqnarray}
Completeness formula:
\begin{equation}
1=\int d{\bar \xi}d\xi d{\bar \xi}_0d\xi_0|\xi,\xi_0\rangle e^{-{\bar \xi}^c\xi^b-{\bar \xi}^b\xi^c-{\bar \xi}_0\xi_0}\langle\xi,\xi_0|.
\end{equation}

\end{document}